\documentclass[12pt]{amsart}

\usepackage{amsfonts,amsmath,amssymb,amsthm,float}

\newcommand{\ra}{\rightarrow}

\newcommand{\RR}{{\mathbb R}}
\newcommand{\CC}{{\mathbb C}}
\newcommand{\DD}{{\mathbb D}}

\newcommand{\Spec}{{\Phi{\rm Spec}}}
\newcommand{\id}{{\rm id}}

\newtheorem{theorem}{Theorem}[section]

\newtheorem{lemma}{Lemma}[section]

\newtheorem{defn}{Definition}[section]

\begin{document}

\author{Florin Moldoveanu}
\address{Logic and Philosophy of Science Research Group, University of Maryland at College Park}
\title{Non viability of hyperbolic quantum mechanics as a theory of Nature}

\begin{abstract} 
Quantum and classical mechanics share a common algebraic formalism which is expressed naturally in the language of category theory. A third realization of this formalism is the so-called hyperbolic quantum mechanics where split-complex numbers replace the usual complex numbers. We introduce and explore the corresponding generalization of C*-algebras and prove that hyperbolic quantum mechanics is not a viable candidate for describing Nature. Quantum and classical mechanics are the only acceptable theories of Nature which are invariant under tensorial composition.  
\end{abstract}


\maketitle


\section{Introduction}
Quantum and classical mechanics have very similar algebraic mathematical structures centered on observables which play a dual role as observables and generators. In the quantum case one encounters a Jordan-Lie algebra and the corresponding classical mechanics mathematical structure is a Poisson algebra \cite{LandsmanBook}. It is natural to expect that common mathematical structures of classical and quantum mechanics capture essential features of Nature and could point towards possible generalizations. Beside the Jordan-Lie and Poisson algebra formalism, there is another set of axioms introduced by Segal which are obeyed by both quantum and classical mechanics \cite{SegalAxioms}. However, this set of axioms is too general, because Segals' axioms do not demand the algebra to be involutive. It is the involution property of the C*-algebra formulation of quantum mechanics which generates a ``dynamic correspondence'' between observables and generators \cite{AlfsenShultz} and constrains the theory into the Hamiltonian formalism.

The properties of the Jordan-Lie algebra can be trivially checked using the anti-commutator for the Jordan algebra and the commutator for the Lie algebra. It is remarkable that if one demands the invariance of the laws of Nature under tensor composition, the Jordan-Lie and the Poisson algebra structures are obtained with the minimal assumption of the existence of a Lie algebra and its distributive property over the secondary product \cite{GP}.

Physically the invariance of the laws of Nature under tensor composition means that the laws of Nature are the same regardless of how we partition in our mind a physical system into subsystems. Recently this approach was put into a category theory formalism \cite{AKapustin}. 

One can understand quantum and classical mechanics as unique ``fixed points'' mathematical structures under composability. This is why the formalism of quantum mechanics is extremely rigid and not amenable to changes. Quantum and classical mechanics belong to distinct composability classes which one can call elliptic and parabolic composability. The composability class classification comes from the associator identity relating the symmetric product (Jordan product for quantum mechanics or regular function multiplication for classical mechanics) with the skew-symmetric product (the commutator for quantum mechanics and the Poisson bracket for classical mechanics).

There is however a third possibility called a hyperbolic composability class. This gives rise to what is called ``hyperbolic quantum mechanics'' \cite{KhrennikovSegre} which is a quantum mechanics-like theory over split-complex numbers. The viability of hyperbolic quantum mechanics as a theory of Nature (maybe in some limiting case) is a very interesting problem (one speculation is that hyperbolic quantum mechanics may be useful in describing the physics of the interior of a black hole \cite{GrginBook}). Comparison with complex quantum mechanics can also illuminate known mathematical properties of the latter.

Prior attempts \cite{KhrennikovSegre, AKapustin} were made to prove that  hyperbolic quantum mechanics is non-physical, but they relied on additional not fully justified assumptions. The major technical problem is that since hyperbolic quantum mechanics is defined over split-complex numbers, the usual C*-algebraic formalism results are not applicable because C*-algebras are defined over complex numbers. Generalization of C*-algebras to Hilbert modules is not helpful either for the same reason.

When looking at the definition of operator algebras like C*-algebras and JB-algebras \cite{AlfsenShultz}, one notices separate algebraic and norm properties. At first sight they look to have very different origins, but for C*-algebras the norm is unique \cite{ConnesBook} and defined by the spectral radius - an algebraic concept. 

\begin{eqnarray}\label{SpectralRadius}
{||T||}^2 &=& {\rm Spectral~radius~of~} T^* T \\ \nonumber
&=& \max \left\{|\lambda | ; \lambda \in \CC {\rm ~and~} (T^* T - \lambda 1) {\rm ~is~not~invertible}\right\}
\end{eqnarray}
 
Because the algebraic properties are constrained by composability, it is conceivable that the analytic properties are constrained as well. We will show that the usual functional analysis has a natural relevant extension in the split-complex case. We will also introduce the phase space formulation of hyperbolic quantum mechanics and show that hyperbolic Wigner functions can generate overall negative probability predictions. Hence hyperbolic quantum mechanics cannot describe Nature. Because classical mechanics is ruled out by experimental evidence \cite{AspectExperiment}, Nature is quantum mechanical at core, and no generalization is possible.    

\section{The algebraic approach to quantum mechanics}
In the algebraic approach to quantum mechanics one avoids starting with a specific Hilbert space and the primary objects are the fields (or observables) \cite{EmchBook}. The Hilbert space comes later via the Gelfand-Naimark-Segal (GNS) construction \cite{GNSReference} and it is only one possible realization of quantum mechanics. Both quantum and classical mechanics have Hilbert space realizations \cite{ClassicalInHilbertSpace} and similarly, both have phase space formulations \cite{WignerFunctions}. 

Let us start with the essential definitions.

{\defn A Jordan algebra is a symmetric power associative algebra:
\begin{eqnarray*}
A \circ B = B \circ A \\
A \circ (B \circ A^2) = (A \circ B) \circ A^2 
\end{eqnarray*} 
}

\noindent Note that we do not demand the reality condition: $A^2 + B^2 = 0 \Rightarrow A=B=0$.
{\defn A Lie algebra is a skew-symmetric algebra obeying the Jacobi identity:
\begin{eqnarray*}
[ A , B ] = - [ B , A ] \\ \nonumber
[ A , [ B , C ] ] + [ C , [ A , B ] ] + [ B , [ C , A ] ] = 0
\end{eqnarray*}
}

{\defn A product $\alpha$ is called a derivation with respect to a product $\sigma$ if it obeys the Leibniz rule:
\begin{eqnarray*}
A\alpha (B \sigma C) = (A\alpha B) \sigma C + B \sigma (A \alpha C)
\end{eqnarray*}}

{\defn An associator for a product $\alpha$ quantifies the violation of associativity and is defined as:
\begin{equation*}
{[A, B, C]}_{\alpha} = (A \alpha B ) \alpha C - A \alpha (B \alpha C)
\end{equation*}}

{\defn A composability algebra is a real vector space $\mathfrak{A}_{\RR}$ equipped with two bilinear maps $\sigma$ and $\alpha$ such that the following conditions apply:
\begin{eqnarray*}
\alpha {\rm ~is~a~Lie~algebra}\\
\sigma {\rm ~is~a~Jordan~algebra}\\
\alpha {\rm ~is~a~derivation~for~}\sigma {\rm ~and~} \alpha \\
{[ A, B, C] }_{\sigma} + \frac{1}{4} \hbar^2 {[A, B, C]}_{J\alpha} = 0
\end{eqnarray*}
where $J^2 = -1,0,+1$.
} 

\noindent When $J^2 = -1$ the composability algebra is called a Jordan-Lie algebra and when $J^2 = 0$ it gives rise to a Poisson algebra. $J^2 = +1$ corresponds to hyperbolic quantum mechanics over split-complex numbers.

But why do we call this algebra a composability algebra? Suppose we start with a two product algebra and have two physical systems $A$ and $B$. Suppose the products $\alpha_A, \sigma_A$ (one skew-symmetric and one symmetric) apply to system $A$, and correspondingly $\alpha_B, \sigma_B$ apply to system $B$. By tensor composition and symmetry property, the total system $T = A\otimes B$ is described by the following products:
\begin{eqnarray*}
(f_1 \otimes f_2) \alpha_{12} (g_1 \otimes g_2) = a (f_1 \alpha_1 g_1)\otimes (f_2 \sigma_2 g_2) + b (f_1 \sigma_1 g_1)\otimes (f_2 \alpha_2 g_2) \\
(f_1 \otimes f_2) \sigma_{12} (g_1 \otimes g_2) = c (f_1 \sigma_1 g_1)\otimes (f_2 \sigma_2 g_2) + d (f_1 \alpha_1 g_1)\otimes (f_2 \alpha_2 g_2)
\end{eqnarray*}  

\noindent A remarkable result \cite{GP} is that demanding the invariance of the dynamical laws of Nature under tensor composition fixes the parameters $a,b,c,d$ to:
\begin{eqnarray*}
a = b = c = 1\\
d = \frac{J^2 \hbar^2}{4}
\end{eqnarray*}

\noindent In shorthand notation:
\begin{eqnarray}\label{composability}
\alpha_{12} &=& \alpha_1 \sigma_2 + \sigma_1 \alpha_2\\ \nonumber
\sigma_{12} &=& \sigma_1 \sigma_2 + \frac{J^2 \hbar^2}{4} \alpha_1 \alpha_2
\end{eqnarray}

\noindent Moreover, one needs only to demand the existence of a Lie algebra (the product $\alpha$) and its distributive property over the secondary product to get the Jordan algebra property and the associator identity.

Another way to understand the invariance of the dynamic under composability is the uniqueness of the Plank constant \cite{SahooPlank}. 

From the associator identity one can introduce a new product $\beta = \sigma \pm \frac{J\hbar}{2} \alpha$ and we have the following result:

{\lemma The product $\beta = \sigma \pm \frac{J\hbar}{2} \alpha$ is associative.
}
\begin{proof}
In the associator property of the composability algebra, each product appears twice and the relationship is not linear. To prove associativity need to show that cross terms cancel.
Let us compute the associator ${[A, B, C]}_{\beta} = A \beta ( B \beta C) - (A \beta B) \beta C$ using the definition of $\beta$:
\begin{eqnarray*}
{[A, B, C]}_{\beta} = A \beta (B \sigma C \pm \frac{J \hbar}{2} B \alpha C)
- (A \sigma B \pm \frac{J \hbar}{2} A \alpha B) \beta C \\ \nonumber
= A \sigma (B \sigma C) \pm \frac{J \hbar}{2} A \sigma (B \alpha C) \pm \frac{J \hbar}{2} A \alpha (B \sigma C) + \frac{J^2 \hbar^2}{4} A \alpha (B \alpha C) \\ \nonumber
 -(A \sigma B) \sigma C \mp \frac{J \hbar}{2} (A \sigma B ) \alpha C \mp \frac{J \hbar}{2} (A \alpha B) \sigma C - \frac{J^2 \hbar^2}{4} (A \alpha B) \alpha C \\ \nonumber
= {[A, B, C]}_{\sigma} + \frac{\hbar^2}{4} {[A, B, C]}_{J \alpha} \\ \nonumber
\pm \frac{J \hbar}{2} \{ A \sigma (B \alpha C) + A\alpha (B \sigma C) - (A \sigma B) \alpha C - (A\alpha B) \sigma C \} = 0
\end{eqnarray*}

In the last line the terms cancel after using the Leibniz rule for $A\alpha (B\sigma C)$ and $(A \sigma B) \alpha C$.
\end{proof}

The associativity of the product $\beta$ allows for the creation of a state space and the introduction of probabilities. Physically it implies that the outcome of parallel and sequential combination of experiments are blind to how we partition the composite experiment into sub-experiments \cite{GoyalQM}. 

\subsection{Standard representation}

It is easy to check that a realization of Eq.~\ref{composability} is given by the usual commutator and the Jordan product:

\begin{eqnarray*}
A\alpha B &=& \frac{J}{\hbar}(AB - BA)\\ \nonumber
A\sigma B &=& \frac{1}{2} (AB + BA)
\end{eqnarray*}

\noindent In the standard representation the product $\beta$ becomes:
\begin{eqnarray*}
&\sigma -\frac{J \hbar}{2} \alpha &{\rm ~~for~~}J^2 = -1\\ \nonumber
&\sigma +\frac{J \hbar}{2} \alpha &{\rm ~~for~~}J^2 = +1\\ 
\end{eqnarray*}
which is the usual complex or split-complex number multiplication (the other two choices correspond to a reversed order of multiplication).
 
The case of $J^2 = 0$ is that of classical mechanics. In this case the $\alpha$ product becomes the Poisson bracket and it can be shown that the Poisson bracket is the most general way to construct a skew-symmetric product for a commutative associative algebra using the standard argument that two biderivations are equal as soon as they agree on a systems of generators \cite{PoissonBook}.

Care must be exercised in the classical case because $J^2 = 0$ does not imply that $J=0$. Usually it is incorrectly stated that quantum mechanics becomes classical mechanics in the limit $\hbar \rightarrow 0$. The proper limit is for $\hbar$ to become a nilpotent element. 

The space of the product $\beta$ is graded ($\beta \ne \sigma$ even for classical mechanics) and $J$ is in general an involution map between observables  and generators (when $J^2 \ne 0$) or between canonical coordinates (when $J^2 = 0$). 

\subsection{Phase space representation}

From the Poisson bracket, let us define an operator $\overleftrightarrow{\nabla}$ as follows:

\begin{equation*}
\overleftrightarrow{\nabla} = \sum_{i=1}^{N} [ \overleftarrow{\frac{\partial}{\partial x_i}}\overrightarrow{\frac{\partial}{\partial p_i}} - \overleftarrow{\frac{\partial}{\partial p_i}}\overrightarrow{\frac{\partial}{\partial x_i}}]
\end{equation*}

\noindent In the classical case, the composability algebra gives rise to a Poisson algebra with $\alpha = \overleftrightarrow{\nabla} = \{ \cdot , \cdot \}$ the Poisson bracket and $\sigma$ the regular function multiplication.

If in the quantum mechanics case ($J^2 = -1$) we chose the following realization of the products $\alpha$ and $\sigma$:

\begin{eqnarray*}
\alpha &=& \frac{2}{\hbar}\sin ( \frac{\hbar}{2}\overleftrightarrow{\nabla}) \\ \nonumber
\sigma &=& \frac{2}{\hbar}\cos (\frac{\hbar}{2}\overleftrightarrow{\nabla}) 
\end{eqnarray*}

\noindent this satisfies Eq.~\ref{composability} by trigonometric identities and we obtain the phase-space formulation of quantum mechanics where the product $\alpha$ is known as the Moyal bracket \cite{MoyalBracket}. 

In this formulation the associative product $\beta = \sigma + \frac{J \hbar}{2} \alpha$ is called ``the start product'' $\star$:
\begin{equation*}
f \star g = f \sigma g + \frac{J\hbar}{2} f \alpha g = f e^{\frac{J\hbar}{2} \overleftrightarrow{\nabla}} g
\end{equation*}

\noindent If we start from a classical system, a natural question to ask is the equivalence of two star products. Because the star products can be decomposed as an infinite sum of morphisms proportional with a given power of the Planck constant, there could be inequivalent ways of quantization, and hence quantization is not a functor. 

\subsection{Hyperbolic quantum mechanics}
Inspired by the phase space formulation of quantum mechanics we can introduce the phase space realization of hyperbolic quantum mechanics by using the following products:

\begin{eqnarray*}
\alpha &=& \frac{2}{\hbar}\sinh ( \frac{\hbar}{2}\overleftrightarrow{\nabla}) \\ \nonumber
\sigma &=& \frac{2}{\hbar}\cosh (\frac{\hbar}{2}\overleftrightarrow{\nabla}) 
\end{eqnarray*}

\noindent Similarly we can introduce a hyperbolic star product as well:

\begin{equation*}
f \star_{h} g = f \sigma g + \frac{J\hbar}{2} f \alpha g = f e^{\frac{J\hbar}{2} \overleftrightarrow{\nabla}} g
\end{equation*}

\noindent with $J^2 = +1$.

The hyperbolic composability case can also be expressed in a formalism similar with the Hilbert space formulation of regular quantum mechanics by replacing the complex numbers with split-complex numbers \cite{KhrennikovSegre}. However, more needs to be changed because the abstract functional analysis spaces involved are not the usual ones and we will introduce them later following a short review of split-complex numbers.

\section{Split complex numbers preliminaries}
Let us review elementary relevant facts about split complex numbers which we will denote as $\DD$. 

Like complex numbers, split complex numbers have an ``imaginary unit'' $j \ne 1$ but with a different property: $j^2 = +1$. This makes $\DD$ an involutive algebra.

If $x$ and $y$ are the real and imaginary components of a split-complex number $z= x + j y$, there are four possible hyperbolic polar form decompositions based on the values of $x$ and $y$:

\begin{eqnarray}\label{polar decomposition}
z = +\rho (\cosh \theta + j \sinh \theta) &{\rm ~if~} x>0 {\rm ~and~} |x| > |y| \\ \nonumber
z = +\rho (\sinh \theta + j \cosh \theta) &{\rm ~if~} y>0 {\rm ~and~} |y| > |x| \\ \nonumber
z = -\rho (\cosh \theta + j \sinh \theta) &{\rm ~if~} x<0 {\rm ~and~} |x| > |y| \\ \nonumber
z = -\rho (\sinh \theta + j \cosh \theta) &{\rm ~if~} y<0 {\rm ~and~} |y| > |x| \\ \nonumber
\end{eqnarray}

\noindent with $\rho$ a positive real number and $\theta$ a real number. We can call $\rho$ the modulus and $\theta$ the hyperbolic phase.

If $\DD^{n\times n}$ denotes the space of all $n \times n$ matrices with split-complex entries, we can define an inner product on $\DD^{n\times n}$ by:
\begin{equation}\label{InnerProduct}
\langle A,B \rangle = {\rm tr}~(A^* B)
\end{equation}
where $\rm tr$ denotes the trace and $*$ denotes the Hermitean conjugate of a matrix. 

Because of the inner product, the hyperbolic quantum mechanics states are defined only up to a hyperbolic phase. Unlike complex quantum mechanics where the elliptic phase is bounded, in hyperbolic quantum mechanics the identification of all points with the same modulus but different hyperbolic phases makes the topology non-Hausdorff. 

We can introduce an indefinite seminorm for split complex numbers using the 
inner product $z^* z$:

\begin{equation}\label{indefinite split complex seminorm}
|| z || = {\rm sign~}(z^*z) \sqrt{|z^* z|}
\end{equation}

Using the polar form decompositions, it is not hard to check that inside each of the four areas separated by the zero norm boundaries a reversed triangle inequality holds:

\begin{equation}
\bigg| || z + w || \bigg| \geq \bigg| || z || \bigg| + \bigg| || w || \bigg|
\end{equation}

\noindent The root cause of triangle inequality for complex numbers is the fact that the cosine function is bounded from above, while the root cause of the reversed triangle inequality is the fact that the hyperbolic cosine function is bounded from below.

\section{Generalized functional analysis: para-spaces}
Although not universally valid, the reversed triangle inequality changes the entire behavior of the usual functional analysis spaces which can be defined over split-complex numbers and new abstract spaces and concepts are required. 

There is a conversion dictionary between the usual functional analysis spaces and proofs and their corresponding hyperbolic counterparts:

\begin{table}[H]
\begin{tabular}[t]{||l|l||}
\hline 
{\em Elliptic} & {\em Hyperbolic} \\
\hline
~~&~~\\
triangle inequality & reversed triangle inequality\\
sup & inf \\
convergent & divergent\\
bounded & unbounded\\
complete & incomplete\\
~~&~~\\
\hline  
\end{tabular}
\caption{Conversion dictionary from regular functional analysis to hyperbolic functional analysis}
\end{table}

Let us introduce the key functional analysis definitions applicable to spaces over split-complex numbers.

\subsection{Para-metric and para-normed spaces}

{\defn A semi para-metric space is a pair $(X,d)$, where $X$ is a set and $d$ is a semi para-distance function on $X$ defined on $X \times X$ such that for all $x,y,z \in X$ we have:

\begin{eqnarray*}
d\in \RR_{+}~~~~~(PM1)\label{(PM1)}\\ 
d(x, y) = 0 {\rm ~if~} x=y~~~~~(PM2)\label{(PM2)}\\ 
d(x,y)=d(y,x)~~~~~(PM3)\label{(PM3)} \\ 
d(x,y) \geq d(x, z) + d(z, y)~~~~~(PM4) \label{(PM4)} \\
\end{eqnarray*}
with $x,y,z$ in $(PM4)$ connectable by a path not crossing any zero distances points.
}
An example of para-metric space is $\DD$ itself.

{\defn A sequence $\left\{ x_n \right\}$ in a metric space $X = (X,d)$ is said to be para-Cauchy if for every $\epsilon > 0$ there is an $N = N(\epsilon)$ such that:
\begin{equation}
d(x_m , x_n) > \epsilon {\rm ~for~every~ } m,n > N
\end{equation}
The space $X$ is said to be para-incomplete if every para-Cauchy sequence in $X$ diverges.
}

{\theorem Every divergent sequence in a para-metric space $X$ is a para-Cauchy sequence.}
\begin{proof}
If $x_n, x\in X$ and for every $\epsilon > 0$ there is an $N = N(\epsilon )$ such that $d(x_n - x) > \epsilon /2$ for all $n > N$, by reversed triangle inequality:
\begin{equation*}
d(x_m , x_n) \geq d(x_m , x) + d (x, x_n) > \frac{\epsilon}{2} + \frac{\epsilon}{2} = \epsilon
\end{equation*}
for all $m, n > N$. Hence the sequence $(x_n)$ is para-Cauchy.
\end{proof}

Ordinary Cauchy sequences also play a role but with the caveat that converging  sequences are no longer guaranteed a unique limit. To build an intuition about para-Cauchy sequences and para-incompleteness, when the reversed triangle inequality holds one thinks not of (elliptic) boundary value problems, but of (hyperbolic) initial value problems and preservation of causality. Physically it is desirable to shield the local value by the influence of far away points when the topology is non-Hausdorff. 

{\defn An indefinite para seminormed space is a vector space $X$ over split complex numbers $\DD$ with a (not necessarily positive) real-valued function ${||x||}_{X}$ for all $x \in X$ obeying the following properties:
\begin{eqnarray*}
{||\alpha x||}_{X} = {||\alpha||}_{\DD} {||x||}_{X} ~~~~~(PN1)\label{(PN1)}\\
\bigg| ||x+y|| \bigg| \geq \bigg| ||x|| \bigg| + \bigg| ||y|| \bigg| ~~~~~(PN2)\label{(PN2)}
\end{eqnarray*}

\noindent where $\alpha \in \DD$ and with $x,y$ in $(PN2)$ connectable by a path not crossing any zero (para) norm points.
}

\noindent Given any two para-normed spaces, we can consider linear maps between them. In particular, functionals are linear maps to $\DD$ which leads us to the concept of dual spaces. Unlike regular functional analysis, the interesting cases here are the unbounded maps defined as follows:

{\defn Let $X$ and $Y$ be para-normed spaces and $T:\mathcal{D} (T) \rightarrow Y$ a linear operator, where $\mathcal{D}(T) \subset X$. The operator $T$ is said to be unbounded if there is a positive real number $c$ such that for all $x\in \mathcal {D}(T)$
\begin{equation*}
\bigg| ||Tx|| \bigg| \geq c \bigg| ||x|| \bigg|
\end{equation*}
}

\noindent We can also define the corresponding norm of linear operators with the key difference that we are this is defined as infimum and not as supremum:

{\defn The number $||T||$ defined as:
\begin{equation}\label{operator norm}
||T|| = \inf_{\substack{ x\in \mathcal{D} (T)\\ ||x|| \ne 0}} \bigg| \frac{||T x||}{||x||}\bigg| {\rm sign} (||Tx||/||x||)
\end{equation}

\noindent is called the para-norm of operator $T$.
}
 
\noindent Here we note that the condition $||x|| \ne 0$ automatically prevents crossing the boundaries of the domain of the validity of the reversed triangle inequality.

{\lemma An alternative  formula for the para-norm of $T$ is:
\begin{equation}\label{alternative para norm}
||T|| = \inf_{\substack{x\in \mathcal{D} (T)\\ ||x|| = 1}} \bigg| ||Tx||\bigg| {\rm sign} (||Tx||)
\end{equation}
}

\begin{proof}
We write $||x|| = a \ne 0$ and set $y = (1/a) x$ Then $||y|| = 1$ and by linearity:
\begin{equation*}
\bigg | ||T|| \bigg| = \inf_{\substack{ x\in \mathcal{D} (T)\\ ||x|| \ne 0}} \bigg| \frac{1}{a}||Tx||\bigg| = \inf_{\substack{ x\in \mathcal{D} (T)\\ ||x|| \ne 0}} \bigg| \big| \big| T \bigg( \frac{1}{a} x\bigg) \big| \big| \bigg| = \inf_{\substack{ y\in \mathcal{D} (T)\\ ||y|| = 1}} \bigg| ||Ty|| \bigg|
\end{equation*}
\end{proof}

\noindent The reason we have chosen the infimum instead of the supremum in the linear operator norm definition is that following theorem holds:

{\theorem The operator para-norm satisfies $(PN1)$ and $(PN2)$.
}
\begin{proof}
$(PN1)$ is obvious. $(PN2)$ follows from:
\begin{eqnarray*}
\inf_{||x|| = 1} \bigg| ||(T_1 + T_2)x|| \bigg| = 
\inf_{||x|| = 1} \bigg| ||T_1 x + T_2 x|| \bigg| \geq \\
\inf_{||x|| = 1} \bigg( \bigg| ||T_1 x|| \bigg| +
\bigg| ||T_2 x|| \bigg| \bigg) \geq 
\inf_{||x|| = 1} \bigg| ||T_1 x|| \bigg| +
\inf_{||x|| = 1} \bigg| ||T_2 x|| \bigg| 
\end{eqnarray*}
\end{proof}

\noindent If we consider the algebraic properties in addition to norm properties we can introduce a para-normed algebra as follows:
{\defn A para-normed algebra $A$ in a para-normed space which is an algebra such that for all $x,y\in A$:
\begin{equation*}
\bigg| ||x y|| \bigg| \geq \bigg| ||x|| \bigg| \bigg| ||y|| \bigg| 
\end{equation*}

\noindent when $||x||*||y|| > 0$.
}

\noindent Considering para-Cauchy behavior we can restrict the prior definition:

{\defn A para-Banach algebra $A$ is a para-normed algebra which is para-incomplete.}

\noindent Another key theorem regarding para-normed algebras holds:
 
{\theorem The linear operator algebra between two para-normed spaces is a para-normed algebra.
}
\begin{proof}
From Eq.~\ref{operator norm} we have:
\begin{equation*}
\bigg| ||A v|| \bigg|  \geq \bigg| ||A|| \bigg| \bigg| ||v||\bigg| 
\end{equation*}
Applying it twice: 
\begin{equation*}
\bigg| ||AB v|| \bigg| \geq \bigg| ||A|| \bigg| \bigg| ||Bv|| \bigg| \geq \bigg| ||A|| \bigg| \bigg| ||B|| \bigg| \bigg| ||v|| \bigg|
\end{equation*}
and using Eq.~\ref{alternative para norm} we obtain:
\begin{equation*}
\bigg| ||A B|| \bigg| \geq \bigg| ||A|| \bigg| \bigg| ||B|| \bigg|
\end{equation*}
Note that demanding the para-norm to be not zero, again prevents crossing the boundaries of the domain of the validity of the reversed triangle inequality. 
\end{proof}

\subsection{Para-inner product spaces}

Now we can introduce the notion of inner product spaces and we will  make sure the definitions apply for indefinite inner products. It is not the non-Hausdorff property or the para-incompleteness which affects the most the corresponding structure of C*-algebra in the hyperbolic case but the indefinite nature of the inner product.

There are a few subtle points we need to highlight. In complex quantum mechanics, given a Hilbert space $\mathcal{H}$, the set of linear operators on $\mathcal{H}$ is a Hilbert space itself $\mathcal{A} = \mathcal{L} (\mathcal{H})$ with the inner product defined by:

\begin{equation}\label{innerProduct2}
\langle A,B \rangle = {\rm tr}~(A^* B)
\end{equation}

The elements of $\mathcal{A}$ are called the observables of the system. This kind of inner product can be defined on any involutive space, but in general it is not guaranteed to be non-degenerate. In the GNS construction \cite{GNSReference}, given a state $\rho$ of the involutive C*-algebra, one constructs a possible degenerate inner product using $\langle A,B \rangle = \rho(A^* B)$ and arrives at $\mathcal{L} (\mathcal{H})$ by constructing a quotient space. In investigating the hyperbolic case we are not concerning ourselves with the degeneracy problem, with the usage of the state $\rho$, or with the trace, and we simply consider the natural inner product first which will be sufficient for our purposes.

We start by defining two essential algebraic identities.
 
{\defn In an involutive space over split-complex numbers, the following algebraic identity called the Polarization Identity holds:
\begin{eqnarray*}
x^* y = \frac{1}{4}[{(x+y)}^* (x+y) - {(x-y)}^* (x-y)] +\\ \nonumber
\frac{j}{4}[{(x+jy)}^* (x+jy) - {(x-jy)}^* (x-jy)]
\end{eqnarray*}
}

{\defn In any involutive space over split-complex numbers, the following algebraic identity called the Parallelogram Identity holds:
\begin{equation*}
{(x+y)}^* (x+y) + {(x-y)}^* (x-y)= 2( x^* x + y^* y )
\end{equation*}
} 

From Eq.~\ref{indefinite split complex seminorm} we see that ${\rm sign} (z^* z) = {\rm sign} (||z||)$ and $z^* z = {\rm sign} (||z||) {||z||}^2$. We can introduce an indefinite inner product as follows:

{\defn In an involution algebra $A$ over split-complex numbers, we can define an indefinite inner product: 
\begin{eqnarray}\label{inner product definition}
<x, y> = x^* y  = \\ \nonumber
\frac{1}{4} [{(x+y)}^* (x+y) - {(x-y)}^* (x-y)] + \\ \nonumber
\frac{j}{4} [{(x+jy)}^* (x+jy) - {(x-jy)}^* (x-jy)]=\\ \nonumber
\frac{1}{4}[{\rm sign} (||x+y||) {||x+y||}^2 - {\rm sign} (||x-y||) {||x-y||}^2] + \\ \nonumber
\frac{j}{4}[{\rm sign} (||x+jy||) {||x+jy||}^2 - {\rm sign} (||x-jy||) {||x-jy||}^2] 
\end{eqnarray}
with the indefinite seminorm given by Eq.~\ref{indefinite split complex seminorm}.
}

Then we have the following theorem:

{\theorem If $x, y \in \DD$ and $||x|| * ||y|| \geq 0$, the following para-Cauchy-Schwarz inequality holds:
\begin{equation*}
|<x,y>| \geq ||x|| ||y||
\end{equation*}
}
\begin{proof}
By direct application of the hyperbolic polar decomposition of Eq.~\ref{polar decomposition} into Eq.~\ref{inner product definition} for the cases where the norms have the same sign.
\end{proof}

It is not clear if the GNS construction generalizes for split-complex quantum mechanics and if there is a corresponding para-Hilbert space in this case, but we can still introduce the definition of para-Hilbert spaces:

{\defn A para-Hilbert space is an indefinite para-inner product space which is para-incomplete.}

\subsection{No-go results for orthogonal decomposition and Riesz representation}

State spaces demand considering convex sets regardless of composability classes. A key result in the elliptic composability case is that given a point $x$ in an inner product space $X$ and a complete not empty convex set $M$, there is a unique point $y \in M$ such that $||x-y||$ is minimal. This result is a prerequisite for subsequent important results like the factorization of Hilbert spaces in orthogonal complements, and for the Riesz representation theorem \cite{KreiszigBook}. 

This result does not hold in hyperbolic functional analysis and this prevents orthogonal decompositions for para-Hilbert spaces and generalization of Riesz representation theorem. Therefore we have a main theorem:

{\theorem Suppose $X$ is an indefinite inner product space with the inner product given by Eq.~\ref{inner product definition} and $M\ne \varnothing$ a complete convex subset. If $\forall x \in X, \exists y \in M$ such that $\delta = \inf_{\bar{y} \in M} || x - \bar{y} || = ||x-y||$, then $y$ is not necessarily unique.
}
\begin{proof}
Suppose there is $y_0 \in M$ such that $||x-y|| = ||x-y_0 || = \delta > 0$. Then ${||y-y_0||}^2 = {||(y-x) -(y_0 -x)||}^2$ and by the parallelogram identity:
\begin{eqnarray*}
{||y-y_0||}^2 {\rm sign}(||y-y_0||) = 2 {||y-x||}^2 + 2 {||y_0 - x||}^2\\
 - {||(y-x) + (y_0 - x)||}^2 {\rm sign}(||(y-x) + (y_0 - x)||) \\
=2 \delta^2 + 2\delta^2 -4 {||\frac{1}{2} (y+y_0) - x||}^2 {\rm sign}(||\frac{1}{2} (y+y_0) - x||)
\end{eqnarray*} 

Since $M$ is convex, $\frac{1}{2}(y+y_0) \in M$ and $||\frac{1}{2} (y+y_0) -x|| \geq \delta$ 

In turn this implies:

${||y-y_0||}^2 {\rm sign}(||y-y_0||) \leq 0$

If the norm is positive definite this would imply that $y_0 = y$, but if the norm is indefinite the uniqueness is no longer a mathematical necessity. 
\end{proof}

The domain of the hyperbolic functional analysis is just as rich as the standard functional analysis. However our aim is not to give a full description of this domain but to understand the key differences which are applicable to quantum mechanics. To this aim we will define the corresponding para C*-algebras and investigate the spectral radius.

\subsection{Born rule does not apply to hyperbolic quantum mechanics}

In complex quantum mechanics an observable has two norms: the operator norm in the space $\mathcal{L}(\mathcal{H})$ generated by the inner product of Eq.~\ref{innerProduct2} and one given by the spectral radius (see Eq.~\ref{SpectralRadius}). We seek to investigate the generalization of C*-algebras in hyperbolic composability and the behavior of the spectral radius.

We start with a definition. 

{\defn A para C*-algebra is a split-complex involutive para Banach space $\mathfrak{A}$ such that for all $A, B \in \mathfrak{A}$ one has:
\begin{eqnarray*}
\bigg| ||AB|| \bigg|  \geq \bigg| ||A|| \bigg| \bigg|  ||B|| \bigg| \\
\bigg| ||A^* A|| \bigg| = ||A||^2
\end{eqnarray*}
}

In the regular C*-algebras over complex numbers it is trivial to check the C* condition $||A^* A|| = ||A||^2$ for bounded operators on a Hilbert space. The converse process of obtaining a Hilbert space and a representation of the algebra elements as bounded operators on that Hilbert space is nontrivial and involves the GNS construction \cite{GNSReference}. 

For the hyperbolic case, obtaining the para C*-condition for unbounded operators on para-Hilbert spaces is trivial as well. We do not seek to generalize the GNS theorem or the spectral theory, but we will consider a simple illuminating computation involving the spectral radius.

In the usual C*-algebras, the spectral radius $r(A)$ of an element $A$ is bounded from above by the norm: $r(A) \leq ||A||$. In a para C*-algebra we expect the spectral radius to be bounded from below by the absolute value of the norm. Let us compute the following:

\begin{equation}\label{simple spectra}
\min \big\{\big| ||\lambda|| \big|; \lambda \in \DD {\rm ~and~} (T - \lambda I) {\rm ~has~no~inverse} \big\}
\end{equation}
for the simplest case of $T = z I$

If $z = z_1 + j z_2$ and $\lambda = \lambda_1 + j\lambda_2$ with $z_1 , z_2 , \lambda_1 , \lambda_2 \in \RR$ we have that $(z-\lambda)$ has no inverse when: $(z_1 - \lambda_1) = \pm (z_2 - \lambda_2)$. If we call $c = z_1 \mp z_2$ we have that $\big| ||\lambda|| \big|$ achieves the minimum for $\lambda = \frac{c}{2} (1 \mp j)$ and $\big| ||\lambda|| \big| = \sqrt{|\lambda_1^2 - \lambda_2^2|} = 0$. This means that the para spectral radius is zero.

If in Eq.~\ref{simple spectra} the operator is the position operator $T=x I$ with $x\in \RR$ then the position spectra contains the zero eigenvalue - contrary to what one obtains in complex quantum mechanics. Hence the Born rule does not apply to hyperbolic quantum mechanics.

\section{Non-viability of hyperbolic quantum mechanics}
There were two prior attempts to eliminate hyperbolic quantum mechanics, but each of them was based on requirements which are either unnecessarily strong or they lack a clear physical justification. 

First, in \cite{KhrennikovSegre} it was proven the inequivalence of the hyperbolic position and momentum representations. As such the Stone-von Neumann theorem does not hold. We have seen that the proper spaces in a Hilbert space-like formulation of hyperbolic quantum mechanics are para spaces and key results from ordinary inner spaces do not hold. Demanding the rule of the addition of probabilities like in complex quantum mechanics (the starting point of \cite{KhrennikovSegre}) and demanding the Stone-von Neumann theorem to hold is not justified. The Stone-von Neumann theorem does not hold for complex quantum mechanics in the case of field theory \cite{EmchBook} and we do not reject the quantum mechanics for this. Also the rule for the addition of probabilities in complex quantum mechanics does not apply to parabolic composability (classical physics) and it is not a good physical ground for rejecting hyperbolic quantum mechanics which belongs to yet another composability class. 

A recent attempt to reject hyperbolic quantum mechanics (in the finite dimensional case only) was made in \cite{AKapustin}. The rejection is based on two axioms related to the spectrum of an observable. The axioms are valid for complex quantum mechanics, but we have seen that for hyperbolic quantum mechanics the spectral radius gives the lower bound and not the upper bound of the operator norm. Implicit in the two additional axioms are the Born rule and the usual spectral theory, which do not apply in this case (in passing we also note the incorrect rejection of classical mechanics \cite{AKapustin} which does have a Hilbert space formulation.) Only Axiom 9 of \cite{AKapustin} has a clear physical justification in terms of ``phantom observables'' and the fact that a repeated measurement in quick succession must yield the same answer. Axiom 8 of \cite{AKapustin}: 

\noindent{\it ``(Physical spectrum of an observable). For any observable $A\in O(S)$ we are given a nonempty  subset of $\RR$ called the physical spectrum of $A$ and denoted $\Spec (A)$. For any polynomial function $f:\RR\ra \RR$ one has $\Spec ( f(A))=f(\Spec(A))$. The physical spectrum of a constant observable $\lambda\cdot \id_S$ is the one-point set $\{\lambda\}$.''}

\noindent demands that the spectrum of an observable function is the function of the spectrum. While valid in complex quantum mechanics, this is a technical requirement with no clear physical justification.

To complete the hyperbolic quantum mechanics rejection as a theory of Nature we should either explore the spectral theory for hyperbolic quantum mechanics and determine if Axiom 8 of \cite{AKapustin} applies or not in this case, or we should find another physically justified criterion for the rejection. 

Because we do have the phase space formalism available for hyperbolic quantum mechanics, we do not need to solve the hard problems of the proper GNS construction and spectral theory generalizations and we will pursue the second option. 

We have seen that the orthogonal decomposition, Riesz representation, and Born rule are not applicable in the hyperbolic case. Each of those results make essential use of positivity and this hints that lack of positivity is what prevents hyperbolic quantum mechanics to be a theory of Nature. Let us prove that this is indeed the case.

John Wheeler famously stated ``it from bit'' \cite{ItFromBit}, but we can paraphrase and state that ``{\em it} is what can generate a bit'' and use this as a criterion for deciding the physical validity of composability algebras. In other words, if a composability algebra is to describe Nature, it must make overall positive probability predictions. In state space, the Wigner functions can be negative and they may be interpreted as negative ``quasi-probabilities'', but overall the predictions are positive.
  
In complex quantum mechanics in phase space formulation, the expectation value of real star squares $g^{*}(x,p) \star g(x,p)$ is always positive even when the probability distribution contains negative parts: $< g^* \star g >~ \geq ~0$

The computation is as follows \cite{TCurtright}:
\begin{eqnarray*}
\int dxdp (g^* \star g) F = (2 \pi \hbar) \int dxdp (g^* \star g) (F \star F) \\
= (2 \pi \hbar) \int dxdp (g^* \star g) \star (F \star F) \\ 
= (2 \pi \hbar) \int dxdp (g^* \star g \star F ) \star F \\ 
= (2 \pi \hbar) \int dxdp F \star (g^* \star g \star F ) \\ 
= (2 \pi \hbar) \int dxdp (F \star g^* ) \star (g \star F ) \\ 
= (2 \pi \hbar) \int dxdp (F \star g^* ) (g \star F ) \\ 
= (2 \pi \hbar) \int dxdp {(g \star F)}^* (g \star F ) \\ 
= (2 \pi \hbar) \int dxdp {|F \star g |}^2 
\end{eqnarray*}
where $F$ is a non necessarily positive Wigner function corresponding to a pure state ($F = (2 \pi \hbar) F \star F$).

The same computation holds in hyperbolic quantum mechanics as well: $< g^{*} \star_h g > = (2 \pi \hbar) \int dxdp {|F \star_h g |}^2 $. The key point in the proof is that the final answer is given as an integral of a number of the form $z^* z$. In complex numbers this is always positive, but not in split-complex numbers. 

\section{Reconstructing quantum mechanics} 
What we have seen so far is that quantum and classical mechanics can be derived from two very general principles: composability and positivity. Composability demands the invariance of the dynamic under tensor composition and positivity demands to be able to generate information about Nature. 

Various attempts were made over time to derive quantum mechanics using those and other ingredients as well. For example, adding the positivity condition to the definition of Jordan algebras makes them ``real Jordan algebras'' and their full classification is well known \cite{JordanAlgebras}. In turn this restricts the allowed number system for quantum mechanics and has a deep connection with projective spaces over the division algebras. 

Exploiting the relationship with projective spaces, Piron was able to reconstruct quantum mechanics \cite{PironQM}. 

On the composability side it is worth mentioning the attempt of Hardy \cite{HardyQM} which uses composition arguments in the context of an instrumentalist derivation, and that of Goyal, Knuth, and Skilling \cite{GoyalQM}. 

The reconstruction project of Masanes and Muller \cite{MasanesQM} is using composition and quantum information requirements.

Other recent attempts \cite{FuchsQM, ChiribellaQM, BarnumWilce, DakicBrukner} are approaching the problem in the spirit of information theory selecting distinctions between classical and quantum information. 

Because the elliptic and parabolic composability classes are well separated, we do not seek to find a quantum separation principle and instead we appeal to the overwhelming experimental evidence in favor of quantum mechanics to reject the parabolic composability. 

\section{Conclusion}
The invariance of the dynamical laws of Nature under tensor composition is a very natural physical principle, and this strongly constraints the formalism. There are only three possible composability classes: elliptic - corresponding to quantum mechanics, parabolic - corresponding to classical mechanics, and hyperbolic - corresponding to ``hyperbolic quantum mechanics''.

We explored hyperbolic quantum mechanics in a Hilbert space like formulation and in a phase space formulation. In the process we uncovered a new mathematical landscape stemming from a reversed triangle inequality. This opens up the possibility to categorify functional analysis and extract the theorems which depend only on the triangle inequality. 

Demanding positive probability predictions rejects hyperbolic quantum mechanics as a theory of Nature and we are left with quantum and classical mechanics as the only acceptable theories of Nature. Experimental evidence overwhelmingly rejects classical mechanics and Nature is quantum mechanical at core with no possibility for generalization.

\section*{acknowledgments}
The author would like to thank Nicolae-Coriolan P\u{a}noiu for valuable comments.

\end{document}